\documentclass{ws-mpla}
\usepackage[super]{cite}
\renewcommand{\thesection}{\arabic{section}}
\begin{document}

\markboth{O.V. Babourova, B.N. Frolov, P.E. Kudlaev}
{Approximate axially symmetric solution of the Weyl--Dirac theory}


\title{\bf APPROXIMATE AXIALLY SYMMETRIC SOLUTION OF THE WEYL--DIRAC THEORY OF 
GRAVITATION AND THE SPIRAL GALACTIC ROTATION PROBLEM}

\author{\footnotesize OLGA V. BABOUROVA}
\address{Institute of Physics, Technology and Information Systems, Moscow Pedagogical State University, M. Pirogovskaya ul. 29\\
Moscow 119992, Russian Federation\\
ov.baburova@mpgu.edu}

\author{BORIS N. FROLOV}

\address{Institute of Physics, Technology and Information Systems, Moscow Pedagogical State University, M. Pirogovskaya ul. 29\\
Moscow 119992, Russian Federation\\
bn.frolov@mpgu.edu}

\author{PAVEL E. KUDLAEV}

\address{Institute of Physics, Technology and Information Systems, Moscow Pedagogical State University, M. Pirogovskaya ul. 29\\
Moscow 119992, Russian Federation\\
pavelkudlaev@gmail.com}

\maketitle


\begin{abstract}
On the basis of the Poincar\'{e}--Weyl gauge theory of gravitation, a new conformal Weyl--Dirac theory of gravitation is proposed, which is the gravitational theory in Cartan--Weyl spacetime with the Dirac scalar field  representing the dark matter model. A static approximate axially symmetric solution of the field equations in vacuum is obtained. On the base of this solution in the Newtonian approximation one considers the problem of rotation  velocities in spiral components of galaxies.  

\end{abstract}

\ccode{PACS Nos.: 95.30.Sf, 04.50.Kd, 95.35.+d, 98.35.Hj}


\section{Introduction}

In Revs. \refcite{BFZ1}--\refcite{BFZ3} the Poincar\'{e}--Weyl gauge theory of gravitation has been developed. On the base of this theory the new conformal Weyl--Dirac theory of gravitation is proposed, which is the gravitational theory in Cartan--Weyl spacetime with the additional geometrical structure as a Dirac scalar field,\cite{Dirac} representing the dark matter model.\cite{BabFr_ArX1,BabFr_ArX2,BFKost}  The post-Riemannian Cartan--Weyl space has geometrical properties of curvature, torsion, and nonmetricity of the Weyl's type, in which a Weyl 1-form is defined by the gradient of the Dirac scalar field $\beta$. The resulting  theory of gravitation has been named by {\it Weyl--Dirac theory of gravitation}, since it generalizes the Einstein--Cartan theory of gravitation to include nonmetricity of the Weyl's type in the Cartan--Weyl space with the additional Dirac scalar field.
 
In Refs. \refcite{BFKudRom}--\refcite{BFKR_ArXive} the spherically symmetric solution of this theory was found, which was conformal to the Yilmas--Rosen metrics of the  Majumdar--Papapetrou class of metrics.\cite{Yilmaz,Rosen,Majum,Papap1} In the given work an approximate axially symmetric solution of the Weyl--Dirac theory of gravitations will be obtained,  and its possible application to the well-known problem of rotation curves of spiral galaxies will also be considered. 

\section{Lagrangian density and field equations 
in the axially symmetric case} 

As a Lagrangian density of the theory we shall take the Lagrangian density $\mathcal L_G$ in vacuum proposed in Refs. \refcite{BabFr_mon}--\refcite{BF_ResInt}, in which the term with the effective cosmological constant and terms square-law on curvature are omitted,     
\begin{eqnarray}
\mathcal L_G &=& 2f_{0} \biggl [ \frac{1}{2} \beta ^{2} \mathcal R^{a}{} _{b} \wedge \eta _{a}{} ^{b} + \rho _{1} \beta ^{2} \mathcal T^{a} \wedge \ast\mathcal T_{a} +\rho _{2} \beta ^{2} (\mathcal T^{a} \wedge \theta _{b} )\wedge \ast(\mathcal T^{b} \wedge \theta _{a} )
\nonumber \\ 
&&+  \rho _{3} \beta ^{2} (\mathcal T^{a} \wedge \theta _{a} )\wedge \ast(\mathcal T^{b} \wedge \theta _{b} )+ 
4\xi \beta ^{2} \mathcal Q_{ab}\wedge \ast\mathcal Q^{ab}+4\zeta \beta ^{2} \mathcal Q_{ab}\wedge \theta ^{a} \wedge \ast\mathcal T^{b}  \nonumber\\
&&+ {\it l}_{1} \rm d\beta\wedge\ast \rm d \beta + {\it l}_{2} \beta \rm d \beta \wedge \theta^{\it a} \wedge \ast\mathcal T_{\it a}  + {\it l}_{3} \beta \rm d \beta\wedge \ast\mathcal Q \frac{}{} \biggr ]\,   \nonumber\\
&& +\beta ^{4} \Lambda ^{\it ab} \wedge \left (\mathcal Q_{\it ab} -\frac{1}{4} g_{\it ab} \mathcal Q \right )\,.\label{LagrPL2}
\end{eqnarray}
Here $\wedge $ is the exterior multiplication symbol, $\rm d $ is the exterior differentiation symbol, $*$ is the Hodge dual conjugation. In Eq. (\ref{LagrPL2}) $\mathcal R^{a}{} _{b}$  and $\mathcal T^{a}$  are the curvature and torsion 2-forms, $\mathcal Q_{ab}$ is the nonmetricity 1-form. The first term here is the Gilbert--Einstein Lagrangian density 4-form generalized to the Cartan--Weyl space ($\eta_a{}^b=\ast(\theta_a\wedge\theta^b$), $f_0 = c^4/16\pi G$). 

We use a variational technique in the formalism of exterior forms on the base of the Lemma on the commutation rule between variation and Hodge star dualization.\cite{BabFr_mon,BFK_CQG} In Eq. (\ref{LagrPL2}) the last term $\Lambda ^{ab}$ contains the 3-form of undetermined Lagrange multipliers, upon variation over which the Weyl condition  $\mathcal Q_{ab} =\frac{1}{4} g_{ab} \mathcal Q$ ($\mathcal Q$ is the Weyl's 1-form of the nonmetricity trace) arises as the field equation. 

The independent variables of the theory are the nonholonomic connection 1-form $\Gamma^{a}{}_{b}$, the cobasis 1-forms $\theta^{a}$, and the Dirac scalar field $\beta$. We obtain three field equations: the $\Gamma$-equation, the $\theta$-equation, and the  $\beta$-equation, which  represent a special case of the field equations received in Refs. \refcite{BabFr_mon}--\refcite{BF_ResInt} (in view of the simplifications accepted at Eq. (\ref{LagrPL2})). 

As well as in the spherically symmetric case,\cite{BFKudRom,BFKR_Vestn-RUDN,BFKR_ArXive} we adopt that the torsion 2-form has the form, 
\begin{equation}
{\mathcal T}^{a}=(1/3)\mathcal T \wedge \theta^{a}\,,\qquad \mathcal T = *(\theta _{a} \wedge *\mathcal T^{a})\,, \label{Tor}
\end{equation}
where $\mathcal T $ is a torsion trace 1-form, and also that a torsion 1-form and a nonmetricity trace 1-form (a Weyl's 1-form) are determined by a gradient of scalar Dirac field
\begin{equation}
\mathcal T = s{\rm d}U \, , \qquad \mathcal Q = q{\rm d}U\,, \qquad   U=\ln \beta\,, \label{TQ}
\end{equation}
where  $s$ and $q$ are arbitrary numbers determined by the coupling constants of the Lagrangian density (\ref{LagrPL2}). 

By analogy to the spherically symmetric case,\cite{BFKudRom,BFKR_Vestn-RUDN,BFKR_ArXive} we search the metrics of the axially symmetric solution of these equations in the form 
\begin{equation}
\label{metr}
ds^2= e^{-2U}\left ( e^{-2\mu}dt^2-e^{2(\mu+\nu )} (d\rho^2+dz^2)+{e^{2\mu}}\rho^2 d\phi^2\right )\,, 
\end{equation}
where $\rho,\;\phi,\;z$ are the axial coordinates. Here $\mu = \mu(r)$, $r^2 = \rho^2 + z^2$, $\nu = \nu(\rho,z)$, $U = U(\rho,z)$, and $c = 1$.

One can obtain the two consequences of the $\Gamma$-equation,
\begin{eqnarray}
&& \frac{2}{3}(\rho_1 - 2\rho_2 - 1)s + \left ( \zeta + \frac{1}{4}\right )q = 2 - l_2 \,, \label{cons1}\\
&& \xi q + \frac{1}{2}\left (\zeta + \frac{1}{4}\right )s - \frac{3}{64}q + 
\frac{1}{2}\left (l_3 + \frac{3}{4}\right ) = 0\,. \label{cons2}
\end{eqnarray}

In analogy with the Refs. \refcite{BFKudRom}--\refcite{BFKR_ArXive}, we choose the unknown function $\mu(r)$ as,
\begin{equation}
\mu(r)  =  \frac{m}{r} = \frac{m}{\sqrt{\rho^2 + z^2}}\,. \label{mu}
\end{equation}
Then the $\theta$-equation  yields the following field equations,
\begin{eqnarray}
&& \nu^{\prime\prime}_{\rho\rho} + \nu^{\prime\prime}_{zz} + \frac{m^2}{r^4}  \nonumber\\
&&- (U^{\prime2}_{\rho} + U^{\prime2}_{z})\left (l_1 + \frac{1}{2}l_2 s + l_3 q - \frac{1}{8}q + \frac{1}{3}s + 1 \right ) = 0\,, \label{ur5}  \\
&& \frac{1}{\rho}\nu^{'}_{\rho} - \frac{m^2 (\rho^2 - z^2 )}{r^6} \nonumber \\
&&+ U^{\prime2}_{\rho} \left (l_1 + \frac{1}{2}l_2 s + l_3 q + \frac{3}{8}q - s - 3 \right )  \nonumber  \\
&&- U^{\prime2}_{z}\left (l_1 + \frac{1}{2}l_2 s + l_3 q - \frac{1}{8}q + \frac{1}{3}s + 1 \right ) = 0\,,  \label{ur6} \\
&& \frac{1}{\rho}\nu^{\prime}_{\rho} - \frac{m^2 (\rho^2 - z^2 )}{r^6} \nonumber \\
&+& U^{\prime2}_{\rho} \left (l_1 + \frac{1}{2}l_2 s + l_3 q - \frac{1}{8}q + \frac{1}{3}s + 1 \right )  \nonumber \\
&&- U^{\prime2}_{z}\left (l_1 + \frac{1}{2}l_2 s + l_3 q + \frac{3}{8}q - s - 3 \right ) = 0\,,  \label{ur7} \\
&& \frac{1}{\rho}\nu^{\prime}_{z} - \frac{2m^2 \rho z}{r^6}\nonumber \\
&&+ 2U^{\prime}_{\rho} U^{\prime}_{z}\left (l_1 + \frac{1}{2}l_2 s + l_3 q + \frac{1}{8}q - \frac{1}{3}s - 1 \right ) = 0\,, \label{ur8}
\end{eqnarray}
and also the $\beta$-equation comes to the form,
\begin{eqnarray} 
&& \nu^{\prime\prime}_{\rho\rho} + \nu^{\prime\prime}_{zz} + \frac{m^2}{r^4}  \nonumber\\
&&- (U^{\prime2}_{\rho} + U^{\prime2}_{z})\left (l_1 + \frac{1}{2}l_2 s + l_3 q + \frac{3}{8}q - s - 3 + \frac{2}{3}\rho_2 s^2 (1 - 2\rho_1 )\right )  \nonumber\\
&&+ \left (U^{\prime\prime}_{\rho\rho} + U^{\prime\prime}_{zz} + \frac{1}{\rho}U^{\prime}_{\rho}\right ) 
\left (l_1 + \frac{1}{2}l_2 s + l_3 q + \frac{3}{8}q - s - 3 \right ) = 0\,. 
 \label{ur9}
\end{eqnarray}

The equations (\ref{ur6}) and (\ref{ur7}) have the consequence, 
\begin{equation}
\frac{3}{8}q - s - 3 = 0\,, \label{qs}
\end{equation}
which is the same condition as in the spherically symmetric case.\cite{BFKudRom,BFKR_Vestn-RUDN,BFKR_ArXive}

If we introduce the condition,
\begin{equation}
\rho_2 s^2 (1 - 2\rho_1 ) = 0\,. \label{sr12}
\end{equation}
then, as its consequence, the field equations (\ref{ur5}) and (\ref{ur9}) yield the equation,
\begin{equation}
U^{\prime\prime}_{\rho\rho}  + \frac{1}{\rho}U^{\prime}_{\rho} + U^{\prime\prime}_{zz} = 0\,, \label{eqU}
\end{equation}
which allows to determine the unknown function $U = U(\rho,z)$.

We find the particular solution of this equation obeying the condition, $U^{\prime\prime}_{zz}=0$,\cite{BFK_Izv-VUZ} 
\begin{equation}
U(\rho,z) = -\frac{m}{R} \left (1 -  \frac{|z|}{h}\right )\ln{\frac{\rho}{R}}\,,\; \quad m\,,R\, = \rm {const}\,,\; \quad \frac{m}{R}\ll 1\,, \;\quad   \frac{R}{\rho} < 1 \,, \label{U}
\end{equation}
where  $m\,,R\,,h$ are arbitrary constants.

\section{Approximate axially symmetric solution and\\ the rotation curves of spiral galaxies}  
  
  Following Ref. \refcite{Kog-Krech}, we shall apply the  axially symmetric solution (to be found) to the well-known problem of the rotation curves of spiral galaxies and interpret this solution as describing the gravitational field of both the spiral component and the bulge spiral galaxies.\cite{BFK_Izv-VUZ}  
  We shall interpret the constants $R$ and $h$ in Eq. (\ref{U}) as radii of galactic bulge and halo, accordingly. The constant $h$ is large enough so that the solution will describe the motion of all stars of the galaxy, for which the condition, $|z|/h < 1$, is fulfilled.  We shall interpret the constant $2m$  in Eq.  (\ref{U}) as the gravitational radius of the bulge with mass $M$, $2m = 2GM/c^2 = r_g$, $\frac{2m}{R}\ll 1$.  
  
We consider the Newtonian approximation of the metrics (\ref{metr}) in case (\ref{U}). As a consequence of the chosen notations, the quantities $U\,, U'_{\rho}\,, U'_{z}$ are small enough in order that quantities $(U'_{\rho})^2\,, (U'_{z})^2$ can be omitted in the equations (\ref{ur5})--(\ref{ur9}) in the first approximation. 
  
As a consequence of this, the field equations Eqs. (\ref{ur5})--(\ref{ur8}) yield the equations,
\begin{equation}
\nu^{''}_{\rho\rho} + \nu^{''}_{zz} + \frac{m^2}{r^4} = 0\,, \qquad 
\frac{1}{\rho}\nu^{'}_{\rho}-\frac{m^2 (\rho^2 - z^2 )}{r^6} = 0\,, \qquad 
\frac{1}{\rho}\nu^{'}_{z} - \frac{2m^2\rho z}{r^6} = 0\,. \label{urnew}
\end{equation}  
The solution of these equations is 
\begin{equation}
\nu(\rho,z) = -\frac{m^2 \rho^2}{2r^4} \,. \label{nu}
\end{equation}  
We see that this quantity is very small.   
  
Therefore we have obtain the approximate axially symmetric solution,\cite{BFK_Izv-VUZ} 
\begin{eqnarray}
&& ds^2= \exp{\left (\frac{2m}{R}\left (1 - \frac{|z|}{h}\right ) \ln{\frac{\rho}{R}} \right ) } ds^2_{GR}\,, \label{appsol}\\
& & ds^2_{GR} =  e^{-\frac{2m}{r}}c^2 dt^2 - e^{\frac{2m}{r}} \left ( e^{\frac{-m^2 \rho^2}{r^4}}(d\rho^2 + dz^2) + \rho^2 d\phi^2 \right )\,. \label{GRsol}
\end{eqnarray}
Here the metrics (\ref{GRsol}) realizes the known axially symmetric solution\footnote{In Ref. \refcite{BFK_Izv-VUZ} an unfortunate misprint appeared in the analogous formula for this metrics.} of General Relativity, which can be found in Ref. \refcite{Singe}. 

In the Newtonian approximation, the metrics (\ref{appsol}) after calculating, 
$g_{00} \approx 1 + \frac{2\varphi}{c^2}$, yields the gravitational potential, 
\begin{equation}
\varphi = \frac{GM}{R} \left (1 -  \frac{|z|}{h}\right )\ln{\frac{\rho}{R}} - \frac{GM}{r} \,. \label{phiN}
\end{equation}
  
Solving the problem of the orbital motion of the stars of the spiral component for this potential, for the square of the velocity of a star we find 
\begin{equation}
v^2 = \frac{GM}{R} \left (1 -  \frac{|z|}{h}\right ) + \frac{GM}{r} \left(\frac{\rho}{r}\right)^2\,. \label{v2}
\end{equation}
 
If the notations introduced before are taken into account, then for stars of the spiral component of the galaxy, the first term in Eq. (\ref{v2}) is significantly greater than the second.  We can conclude that for these stars the condition is valid, $v^2 \approx const$.  Therefore, for any value of $z$ into galaxy arms, the magnitude of the velocity of the orbital motion of a star does not depend substantially on the distance of the star from the center of the galaxy. 

Let us, for instance, take into account the mass structure of Milky Way and its velocity rotation curve. We adopt for the Milky Way Bulge the Hernquist  model,\cite{Hern} and use parameters, $R = 3.5\; pc$, $M = 2*10^{10}M_{\odot}$.\cite{Mbulge} Then put into Eq. (\ref{v2}) $r=R$ and  obtain for the velocity of the star orbital motion the constant magnitude, 
$v\approx 220 \;\frac{km}{s}$, which corresponds with Milky Way observational data. 

The influence of the axial symmetry of the spiral component is decreased at a certain distance from the center, and the influence of the spherical halo of the galaxies becomes predominate, as a consequence of which constancy of the velocities of the stars in the galaxy not to be support. 

It is assumed at the present time that in order to explain the rotation curves of the spiral galaxies the mass of dark matter inside galaxies should be an order of magnitude greater than the mass of the luminous (baryon) component of the galaxies. The solution Eq. (\ref{appsol}) can substantially reduce the required magnitude of the mass of dark matter other then the scalar field in its nature.

\end{document}